\documentclass[12pt]{article}
\pdfoutput=1
\usepackage{amssymb,amsmath,latexsym}
\usepackage{graphicx}
\usepackage{color}
\usepackage[backend=bibtex,sorting=none]{biblatex}

\begin{document}
\title{The sensitivity of the Higgs boson branching ratios to the W boson width}
\author{William Murray, Warwick University / STFC-RAL }
\date{\today}
\maketitle
\abstract{The Higgs boson branching ratio into vector bosons is sensitive
  to the decay widths of those vector bosons  because they are produced with
  at least one 
  boson significantly off-shell. $\Gamma(H\rightarrow VV)$  is approximately
  proportional to the product of the Higgs boson coupling and the
  vector boson width. $\Gamma_Z$  is  well measured, but
  $\Gamma_W$ gives  an uncertainty on $\Gamma(H\rightarrow WW)$ which
  is not negligible. 
  The ratio of branching ratios, BR($H\rightarrow WW$)/BR($H\rightarrow ZZ$)
  measured by a combination of ATLAS and CMS at LHC  is used herein to extract
  a width for the $W$ boson of $\Gamma_W = 1.8^{+0.4}_{-0.3}$~GeV by assuming
  Standard Model  couplings of the  Higgs bosons. This 
%  custodial symmetry. This
  dependence of the branching ratio on $\Gamma_W$ is not discussed in
  most  Higgs boson coupling analyses.

 This is not the work of, nor endorsed by, the ATLAS or CMS collaborations.}

\newpage

\section{Introduction}
The Higgs boson discovered at LHC\cite{CMS:2012nga,Aad:2012tfa} has been the subject of
combined mass\cite{Aad:2015zhl} and couplings\cite{ATLAS-CONF-2015-044}
analyses  by the ATLAS and CMS collaborations. The couplings analysis uses
the so-called $\kappa$ framework of the LHC Higgs cross-section
working group\cite{Dittmaier:2011ti,Heinemeyer:2013tqa}, and relies upon the
cross-section and branching ratio calculations contained therein. This
includes the 
properties of the vector bosons, $W$ and $Z$, for which the  masses
reported in the RPP\cite{PhysRevD.86.010001}, are used to extract
pole masses of $m_Z= 91.15349$ GeV and $m_W = 80.36951$ GeV in
Ref.\cite{Heinemeyer:2013tqa}.  In
addition, 
and especially relevant for this note, the vector boson widths are
calculated   from  their  masses  and assuming the Standard Model(SM), to be
$\Gamma_Z = 2495.81$ MeV and $\Gamma_W= 2088.56$ MeV.
%In other words,
%the  interpretations of the Higgs boson couplings typically assume the
%SM for the vector boson couplings.

The use of the theoretically expected $W$ boson width is not discussed in
Ref.~\cite{Heinemeyer:2013tqa}, it is merely stated. It is not obvious
that this is the best motivated assumption when looking for beyond the
Standard Model  (SM)  effects in Higgs boson
properties. The primary purpose of this document is to highlight that
assumption.

 The widths of the $Z$ and $W$ bosons have also been measured experimentally. 
 The $Z$ boson width is measured via the scan of the $Z$ resonance at
 LEP\cite{ALEPH:2005ab} to be $2495.2\pm2.3$ MeV.
The  $W$ boson width has been measured using mass
reconstruction at LEP 2~\cite{Schael:2013ita}
and with better precision  using  the transverse mass distribution at
the Tevatron~\cite{TEW:2010aj}. These  different 
approaches agree well and are combined in the
RPP\cite{PhysRevD.86.010001} to give $\Gamma_W = 2085\pm42$~MeV,
an error a factor twenty times larger than that for the $Z$.
In consequence,  effects due to the vector boson width uncertainties are
dominated by those from the $W$  boson.

The Higgs boson partial widths and branching ratios  are not
experimentally accessible at 
the LHC, where only products of production and decay can be
studied. However, the ratio of the branching ratios to $WW$ and $ZZ$,
%equivalent to the ratio of the partial widths,
is measurable, and it is presented in
Ref.~\cite{ATLAS-CONF-2015-044}. The measured value of
BR$^{WW}$/BR$^{ZZ}$ is 6.8$^{+1.7}_{-1.3}$.  It is also  accurately
calculable,  using just $m_H$ and the masses and widths  of the $W$,
$Z$ and $H$  bosons. The SM value given
in Ref.~\cite{Heinemeyer:2013tqa}  is 8.09

This ratio is not the only  test of the $H\rightarrow WW$ width 
which could be made. Most obviously the measured rate into  diphotons could be
included in the analysis. However,  further assumptions about the
interaction strengths of all particles entering the decay loop would 
be needed, and there could even be unknown particles.  The analysis
using the vector bosons alone is easier to justify.

\section{Analysis of the widths}

The full calculation  of the Higgs boson partial widths in the SM  is
rather complex. However, the results are tabulated in
Ref.~\cite{Heinemeyer:2013tqa}, and the approach taken here is to use a 
leading-order approximation~\cite{Djouadi:2005gi}, and then scale its
results to those in Ref.~\cite{Heinemeyer:2013tqa} for the nominal
input parameters. This  captures the dependence on the $W$ boson width to a
very good approximation.
The calculation is reproduced below.

\begin{equation}
  \Gamma (H \rightarrow V^* V^* ) = \frac{1}{\pi^2}
  \int_0^{M_{H}^2}\frac{dq_1^2 M_V \Gamma_V}{(q^2_1 - M_V^2 )^2 + M_V^2 \Gamma_V^2}
  \int_0^{(M_{H}-q_1)^2}\frac{dq_2^2 M_V \Gamma_V}{(q^2_2 - M_V^2 )^2
    + M_V^2 \Gamma_V^2} \Gamma_0.
  \label{eq:main}
\end{equation}

In this formula $\Gamma_0$ is

\begin{equation}
  \Gamma_0 = \delta'_V \frac{G_F M_H^3}{16 \sqrt{2} \pi}
  \sqrt{\lambda(q_1^2,q_2^2,M_H^2)} \left( \lambda(q_1^2,q_2^2,M_H^2) +
    \frac{12 q_1^2 q_2^2 }{M_H^4} \right)
\end{equation}

where $\lambda(x,y,z) = (1-x/z - y/z)^2 - 4xy/z^2$ and $\delta'_V$ has
different values depending upon the vector boson:
%$\delta'_W = 1$ and
%$\delta'_Z = 7/12 - 10/9 \sin^2 \theta_W + 40/27 \sin^4 \theta_W$.
$\delta'_W = 2$ and $\delta'_Z = 1$~\cite{Djouadi:2005gi}.
By performing the  integration the partial width can be found.
This calculation assumes the SM coupling strengths to  the $W$ and $Z$
boson.
%In the analysis below only the 2:1 ratio is necessary, and it
%reflects custodial symmetry.

\begin{figure}[th!]
  \begin{center}
          \includegraphics[width=0.45\textwidth]{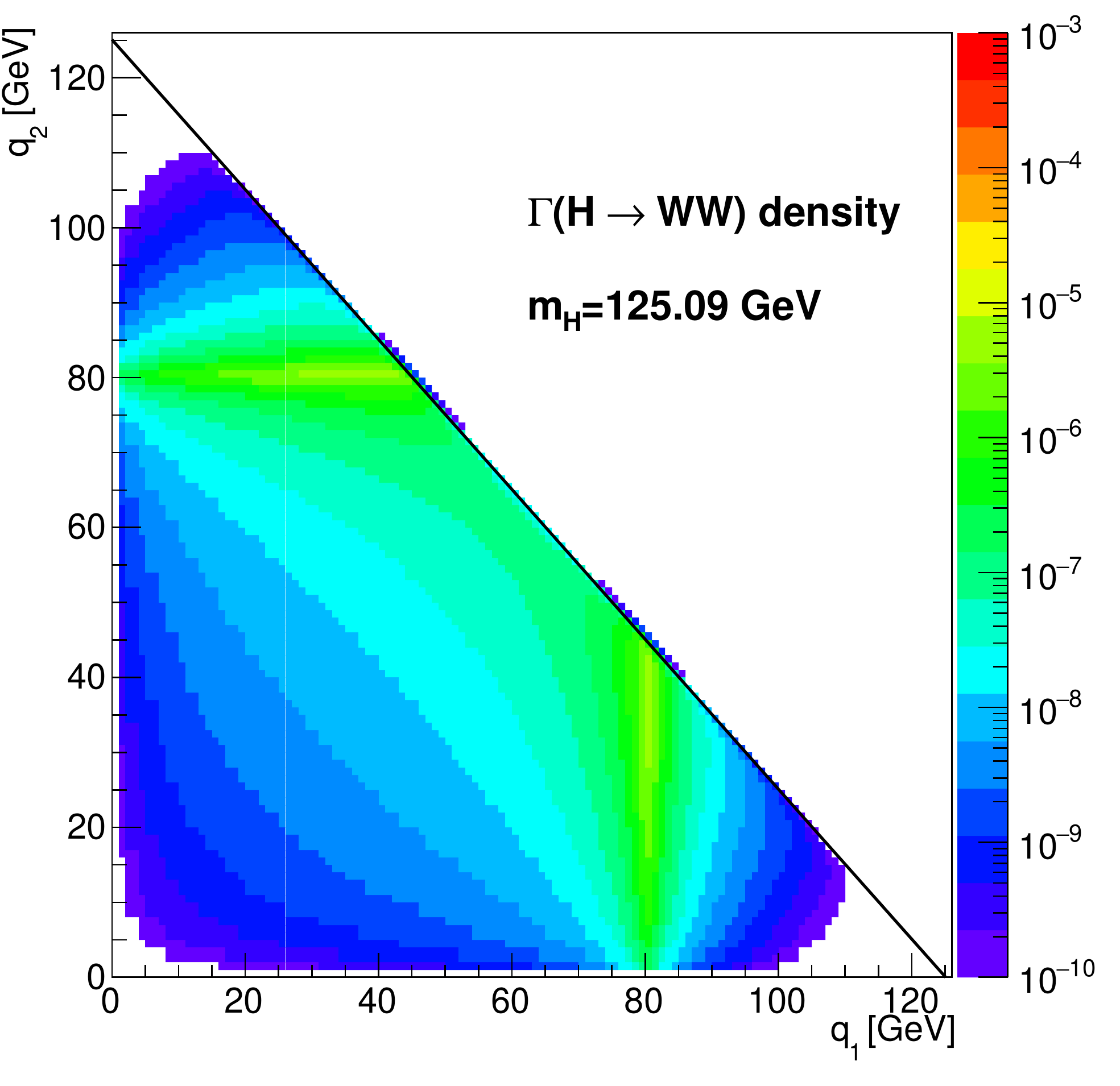}
          \includegraphics[width=0.45\textwidth]{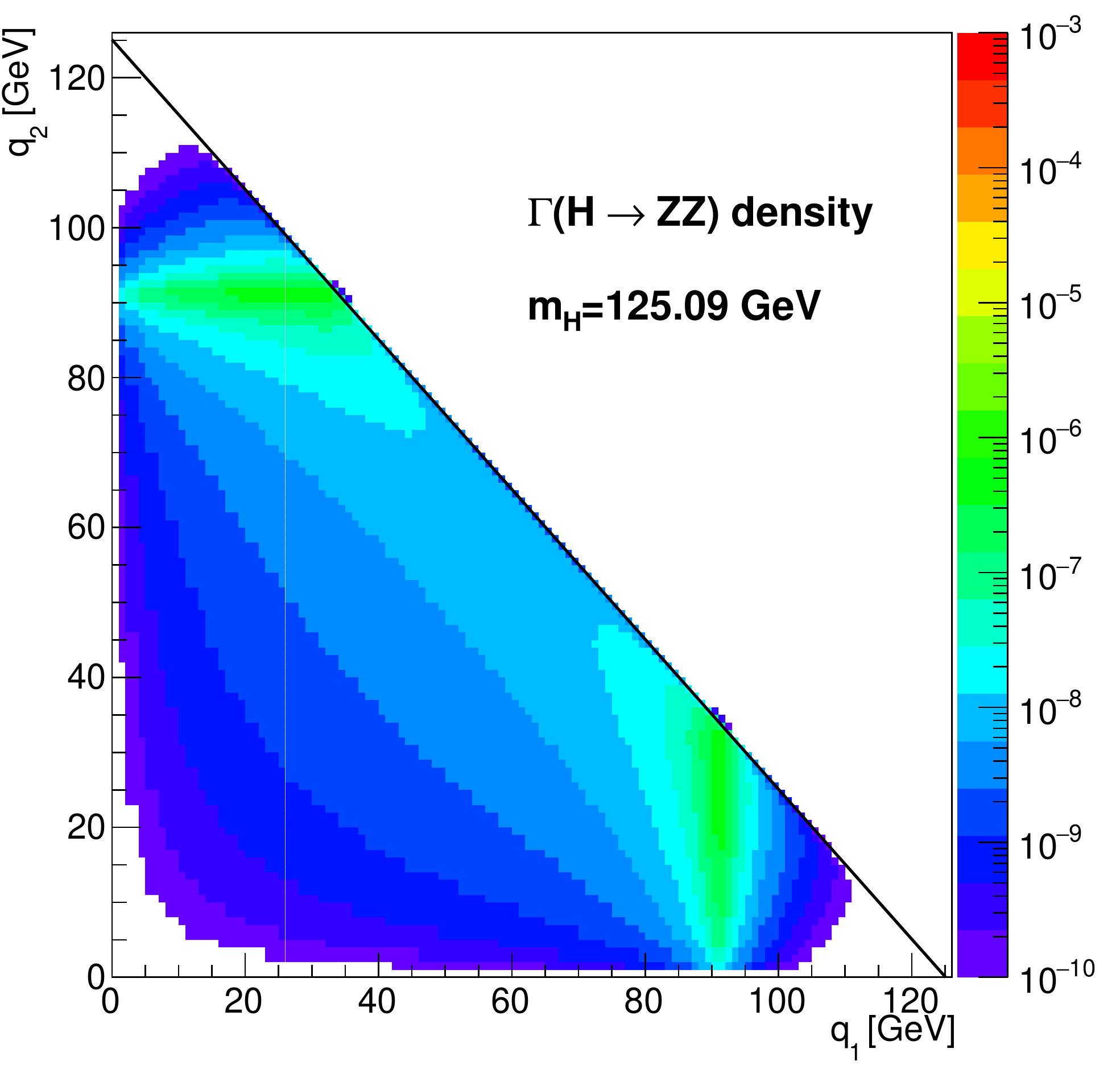}
          \includegraphics[width=0.45\textwidth]{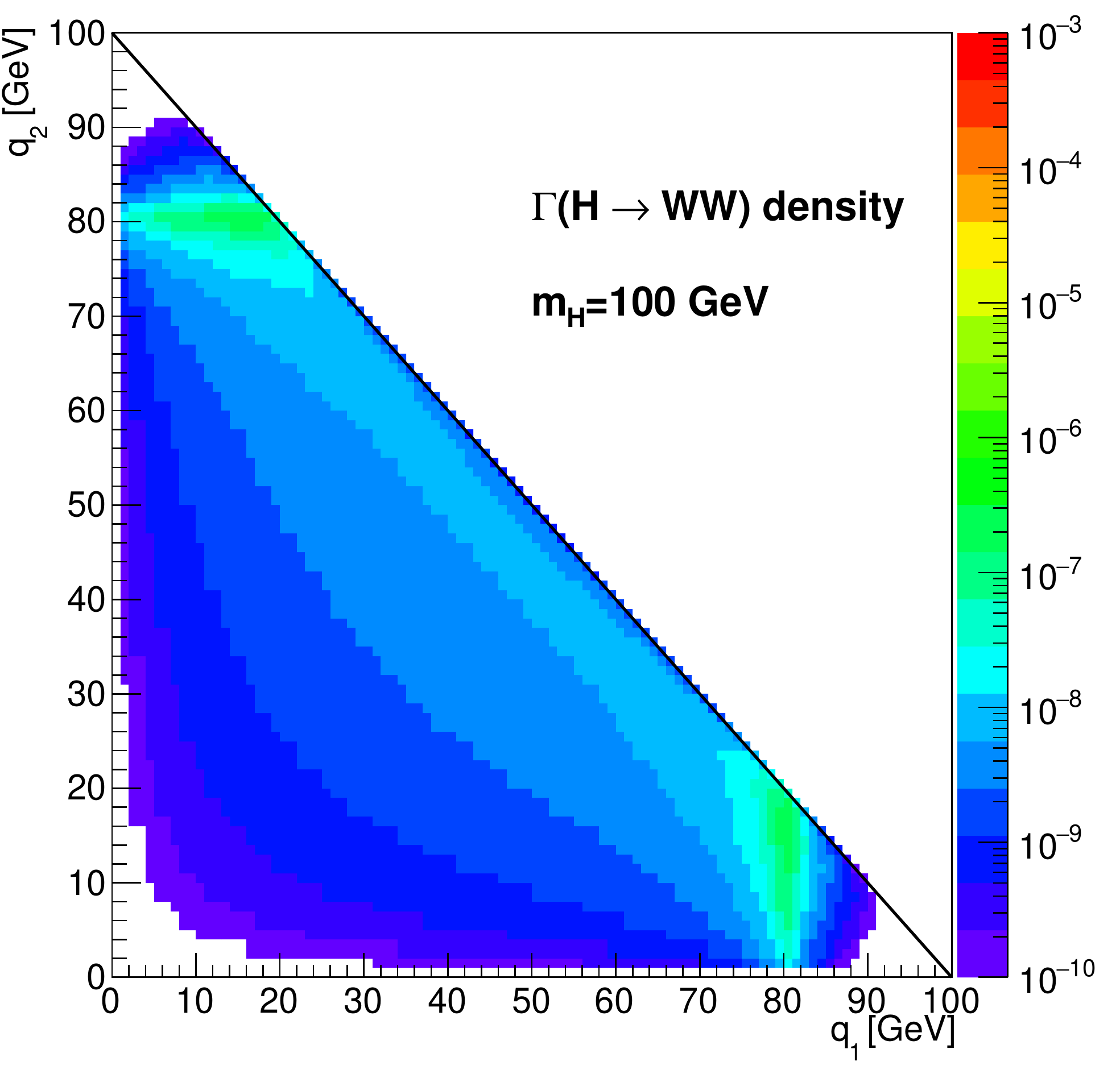}
          \includegraphics[width=0.45\textwidth]{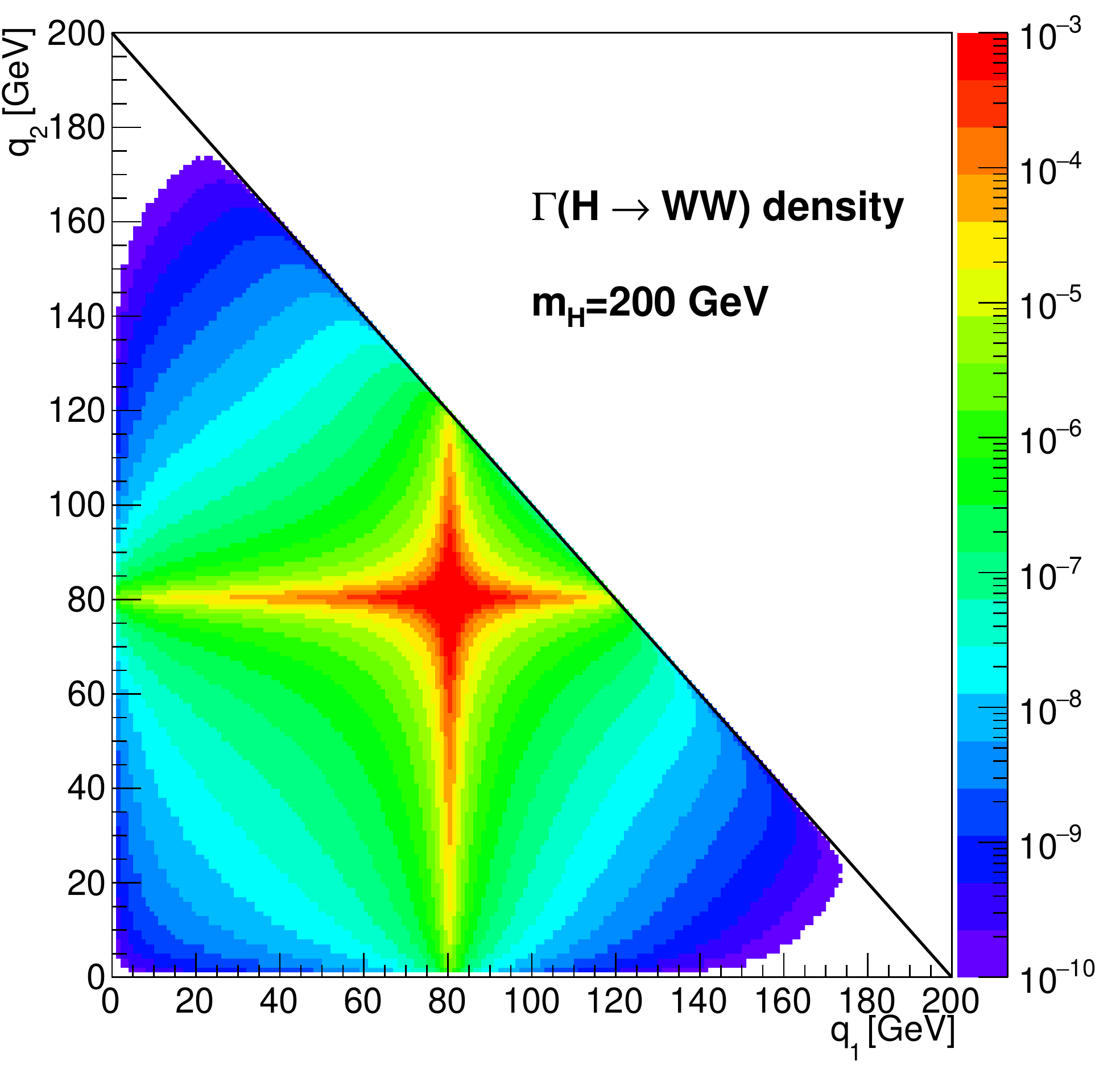}
  \end{center}
  \vspace{-20 pt}
  \caption{The partial width densities of the Higgs boson into various diboson
    mass combinations, in GeV per GeV$^{2}$. Top row has  $WW$ (left) and
    $ZZ$ (right)  for $m_H=125.09$,
    The bottom row shows the $WW$ partial width for $m_H=100$~GeV and
    $m_H=200$~GeV for comparison. The density in this last plot is
    truncated to keep the scale uniformity.}
  \label{fig:content}
\end{figure}

Figure~\ref{fig:content} shows the density of the partial width of the
Higgs to vector boson pairs in the ($q_1,q_2$) plane. It is shown for
Higgs boson masses of 100 and 200 GeV (bottom row) to demonstrate
the impact on the phase space of the Higgs boson mass. For the 200~GeV
case the whole resonant structure is observed, and the factor $\Gamma_V$
in
equation~\ref{eq:main} will largely cancel in the integration. For
lower masses only one, or perhaps no, clear resonant structures dominate
and there will be one 
or two factors of $\Gamma_V$ in the solution.

The numerical evaluation uses the parameters from the LHC Higgs
cross-section working group as given in the introduction and was done
using root~\cite{Brun:1997pa}. To check the
calculation it is first evaluated at  $m_H=126$~GeV because
Ref.~\cite{Heinemeyer:2013tqa} provides  
partial widths at this mass. The values obtained are 0.941 MeV for $WW$ and
0.119~MeV for $ZZ$. These are respectively 97\% and 98\% of the values from
the reference,  0.974~MeV and 0.122~MeV.
%{\color{red} I need to find out more abobut this difference}
This 2-3\% discrepancy with the full calculation shows that the higher
order effects are not large.

Having tested the implementation,  the partial widths are found at
$m_H = 125.09$~GeV. 
%, and a value of $\sin^2 \theta_W$ of 0.2315.
They are  $\Gamma(H\rightarrow WW) = 0.853$ MeV and $\Gamma(H\rightarrow ZZ)$ =
0.107 MeV.

The ratio of the partial widths gives directly the ratio of the
branching ratios, 7.99. This is about 1\% lower then  the 8.09
contained in  Ref.~\cite{Heinemeyer:2013tqa} and the difference is assumed to 
come from the more complete calculation used in that document.
The 2-3\% changes in the WW and ZZ widths have largely cancelled in the ratio.
A scale factor of 1.01
is therefore  applied to subsequent results for
$m_H=125.09$~GeV. This is
reminiscent of the $K$ factor approach to coupling analysis where a
leading order framework is used to calculate a scale factor on  the complete
calculation.

\begin{figure}[th!]
  \begin{center}
          \includegraphics[width=0.45\textwidth]{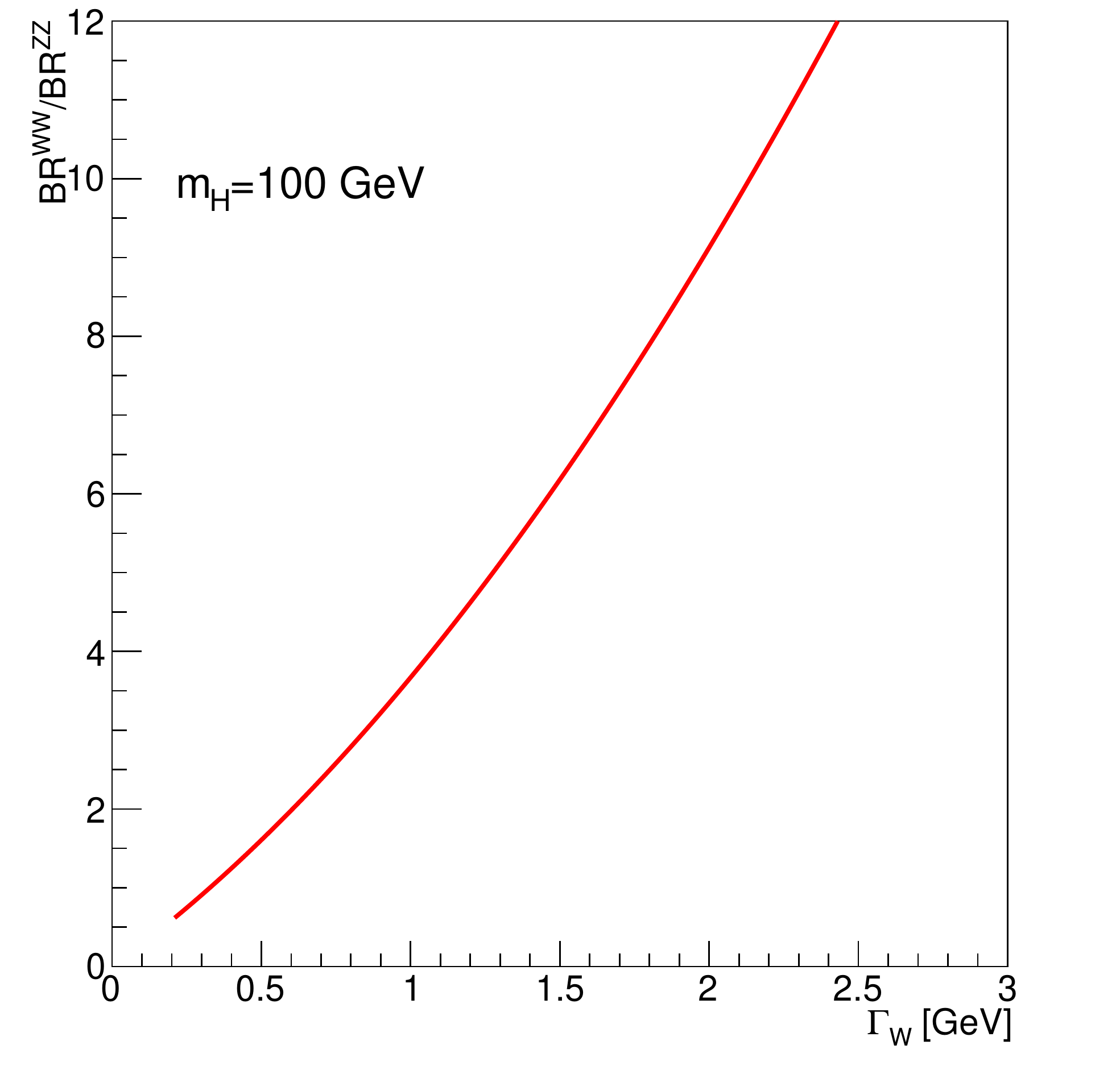}
          \includegraphics[width=0.45\textwidth]{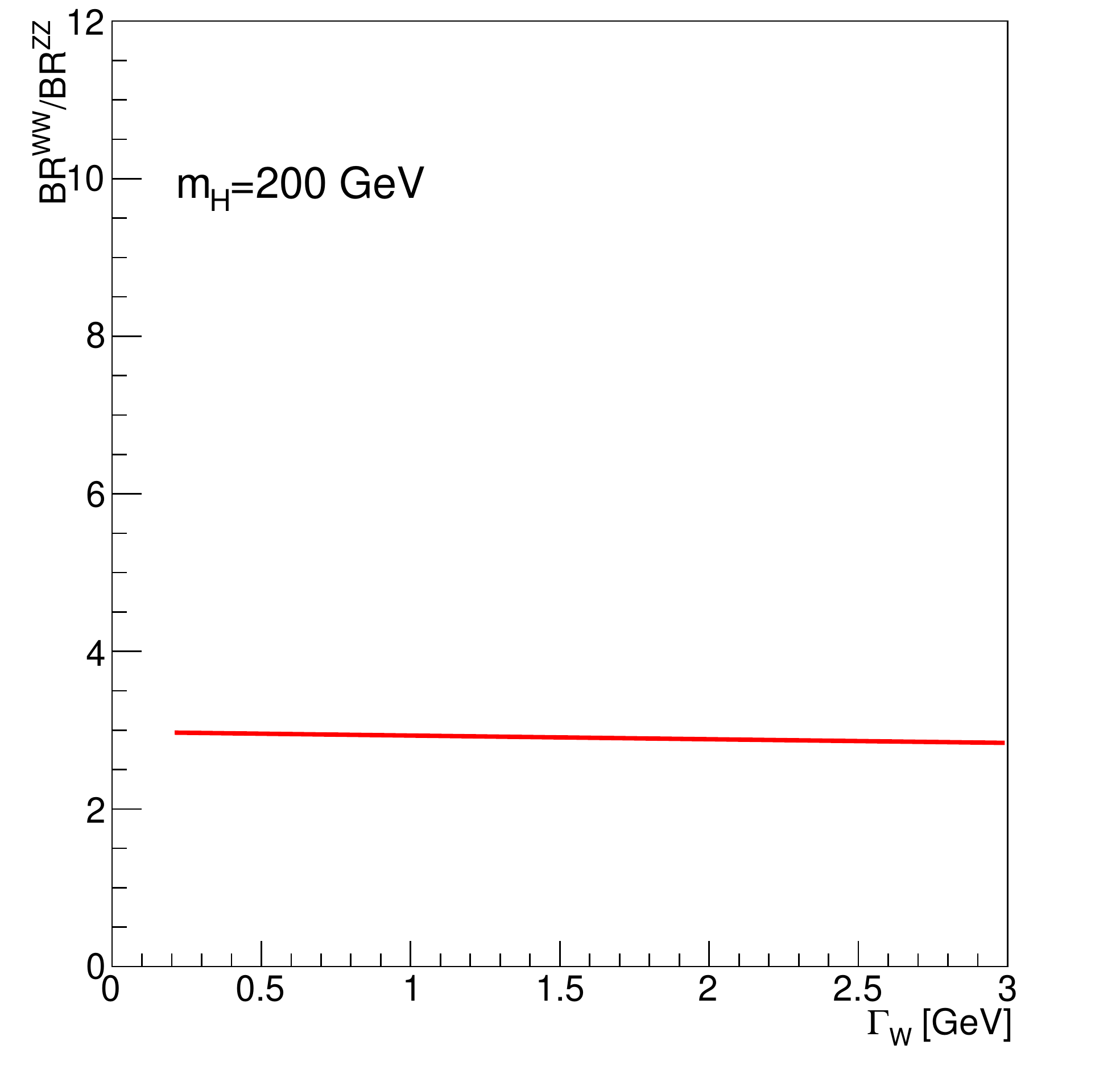}
          \includegraphics[width=0.45\textwidth]{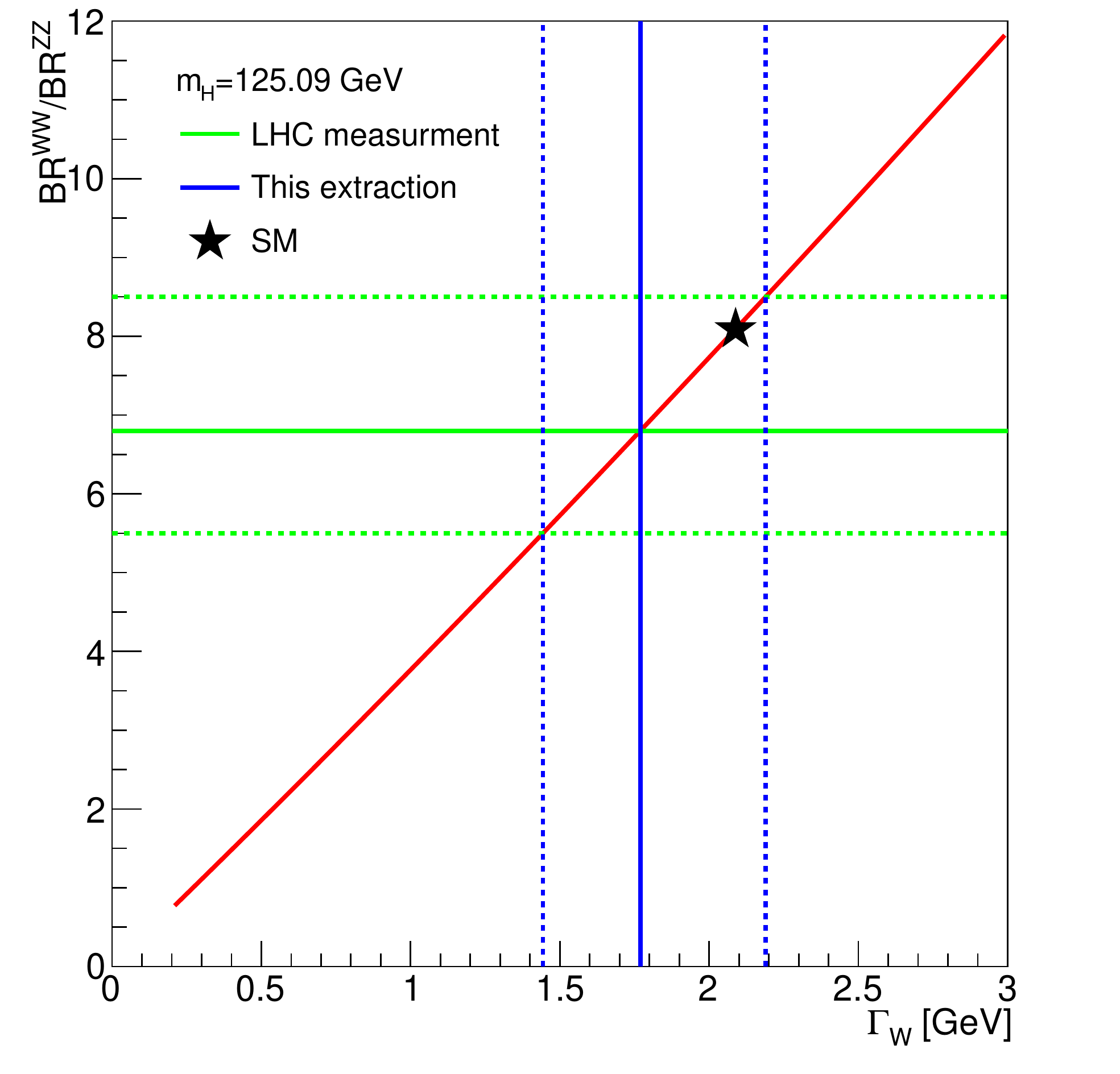}
  \end{center}
  \vspace{-20 pt}
  \caption{The ratio BR$^{WW}$/BR$^{ZZ}$ as a function of the $W$
    boson width, with all other parameters fixed. It is evaluated for
    $m_H$ of 100, 200 and 125.09 GeV in the three subfigures. The
    final figure shows the LHC  measurement of the ratio of Higgs boson branching
    ratios BR($H\rightarrow WW$)/BR($H\rightarrow ZZ$), in green,  the
    extracted $\Gamma_W$ in blue, and the SM expectation is in black.}
  \label{fig:wwid}
\end{figure}

The ratio BR$^{WW}$/BR$^{ZZ}$ as a function of the $W$ width,  ignoring
the uncertainties on all the other parameters, is shown in
figure~\ref{fig:wwid}. Had the Higgs boson decayed to two on-shell
bosons the width would scarcely have entered. If both vector bosons
had been virtual, as is the case for a SM Higgs boson of 100 GeV, the
dependence would have been roughly quadratic. With the actual mass
there is one real and one virtual gauge boson and the width is, to a
good approximation, 
proportional to $\Gamma_W$.  This  supports the  1\% correction via a
 scaling of the ratio to the full calculation.
The equation
is numerically inverted to find the range of widths  which corresponds
to the measured branching ratio range. This is:

\begin{equation}
  \Gamma_W = 1800^{+400}_{-300} \mbox{MeV}
  \end{equation}

An alternative presentation would be to invert the assumptions, and say
that the 2\% uncertainty on $\Gamma_W$ repsrensts a 2\% uncertainty on
$\Gamma(H\rightarrow WW)$ which should be allowed for in the analysis.

\subsection{Errors from the extraction procedure}

The extraction of the ratio
of branching ratios from the LHC data currently has limited precision,
mostly for  statistical  reasons, but also with many systematic
errors. These are not the subject of 
this note, which considers that input as a given. Only the errors
discussed below affect the interpretation.

The Higgs boson mass of $125.09\pm0.21\pm0.11$~GeV has the largest
mass uncertainty in the formula. It changes the extracted value of
$\Gamma_W$ by around 0.2~MeV, which is clearly negligible, and
similarly the $W$ and $Z$  boson masses contribute negligible uncertainty.

The $Z$ boson width is known to 2 per mille, and this translates to a
1 per mille or 2 MeV uncertainty on the prediction of 
$\Gamma(H\rightarrow ZZ)$. This is far
below the precision achievable at LHC and is ignored here. 

The width of the Higgs boson could also influence this result by
changing the relative suppression of $WW$ and $ZZ$ states.
%Upper limits on the  Higgs boson width have been set  by ATLAS and CMS,
%\cite{Khachatryan:2015mma,Aad:2015xua} 
%which, depending upon the asumptions, are less than or of the order
%of 50~MeV. This is considerably less than the uncertainty on  $m_H$
%and its impact has not been explicitly calculated.
%However, these extractions  need assumptions on the Higgs boson
%lineshape.
The tightest model-independent  upper limit on the $H$ boson width is
3.4~GeV from the  CMS studies in the $llll$ final state.\cite{Chatrchyan:2013mxa}
An  integration over the Higgs boson width has not been made, but
its magnitude is estimated by changing the mass by 3.4~GeV, which
gives a 3~MeV shift in the extracted $\Gamma_W$. This is again negligible.

There is a  1\% correction made in the double ratio between the first order
calculation used here and the full calculation.
%If the 1\% shift is interpreted as an error, it  corresponds to
%a change in $\Gamma_W$ of 20~MeV, well below the  experimental
%precision.
However, 
the measured value is compatible with the SM expectation, and so the
calculation has been corrected to the full calculation at least in
some part of the range. The total calculational error is expected to
be dominated by the  uncertainty with which both the $WW$ and $ZZ$
partial widths are 
calculated,  0.5\%\cite{Heinemeyer:2013tqa}. 
A   pessimistic combination of these, 1\%, 
 gives the largest uncertainty on $\Gamma_W$, 20~MeV.

In summary, the total  error of the extraction is estimated to be 20~MeV,
which is negligible in comparison with the experimental error.

\section{Discussion and outlook}

The partial width $\Gamma(H \rightarrow VV)$ is proportional to the
full width of the
vector  boson involved. While it is possible
to impose the SM expectation, this seems to this author a
restrictive way of testing the SM. The alternative, of using the
experimentally measured value. should at least be considered.
The 2\% uncertainty on $\Gamma(H\rightarrow WW)$ from the limited
experimental knowledge of the $W$  boson width is currently well below
to 20\%  uncertainty from the Higgs boson couplings.

The alternative presentation, discussed here, 
treats the Higgs boson physics as known 
and the $W$ as unknown and is perhaps extreme, but under this assumption
$  \Gamma_W = 1800^{+400}_{-300} \mbox{MeV}$ has been extracted.
%The measurements  of the widths of  the  $Z$ and $H$ bosons means they
%do not contribute a significant uncertainty to BR($H\rightarrow
%WW$)/BR($H\rightarrow ZZ)$, and the same is true of all the masses.
A conservative 20~MeV error on the $W$ boson width is estimated due to
uncertainties on the calculation of the partial  widths to $WW$ and $ZZ$.

The uncertainty on this derivation  of $\Gamma_W$ is thus 
dominated by the  errors on the Higgs boson $WW$ and $ZZ$ measurements and will
remain so at HL-LHC.
Various projections for these in the future exist. For example, ATLAS
concluded~\cite{ATL-PHYS-PUB-2013-014} that
5\% and 4\% errors on the $H\rightarrow WW$ and $H\rightarrow ZZ$
signal strength, respectively, were possible using 3000~fb$^{-1}$ if theoretical
systematic errors are ignored. Some of these theoretical errors will
cancel in the ratio, 
so an error approaching  7\% error might be achievable, and presumably
a combination of two experiments  will be better.
At this point a 2\% error on  $\Gamma_W$  would have a
significant impact on the physics interpretation.

\section*{Acknowledgements}

With grateful thanks to Abdelak Djouadi for corrections  and advice.
The author also wishes to acknowledge the financial support of the
STFC of the United Kingdom. 

\printbibliography

\end{document}